\documentclass[letter]{aa} 
\usepackage{txfonts}
\usepackage{graphicx}

\begin{document}

\title{Tidal synchronization of the subdwarf B binary
PG~0101$+$039
\thanks{Based on data from MOST, a Canadian Space Agency mission operated jointly by Dynacon, Inc., the University of Toronto Institute of Aerospace Studies, and the University of British Columbia, with assistance from the University of Vienna}
\thanks{Based on observations collected at the Centro Astron\'omico Hispano Alem\'an (CAHA) at Calar Alto, operated jointly by the Max-Planck Institut f\"ur Astronomie and the Instituto de Astrof\'isica de Andaluc\'ia (CSIC).}
}

\author{S. Geier \inst{1}
   \and S. Nesslinger \inst{1}
   \and U. Heber \inst{1} 
   \and S. K. Randall \inst{2}
   \and H. Edelmann \inst{3} \and E. M. Green \inst{4}}

\offprints{S.\,Geier,\\ \email{geier@sternwarte.uni-erlangen.de}}

\institute{Dr.--Remeis--Sternwarte, Institute for Astronomy, University Erlangen-N\"urnberg, Sternwartstr. 7, 96049 Bamberg, Germany 
   \and ESO, Karl-Schwarzschild-Str. 2, 85748 Garching, Germany
   \and McDonald Observatory, University of Texas at Austin,
     1 University Station, C1402, Austin, TX 78712-0259, USA
   \and Steward Observatory, University of Arizona, Tucson, AZ 85721, USA}

\date{Received \ Accepted}

\abstract{
{\it Aims.} Tidally locked rotation is a frequently applied assumption that 
helps to measure
masses of invisible compact companions in close binaries.  
The calculations of synchronization times are affected by large uncertainties 
in particular for stars with radiative envelopes calling for  
observational constraints. We aim at verifying tidally locked rotation 
for the binary PG~0101$+$039, a subdwarf B star + white dwarf binary 
from its tiny (0.025\%) light variations measured with the MOST satellite
(Randall et al. \cite{randall}).\\  
{\it Methods.} Binary parameters were derived from the mass function, apparent rotation 
and surface gravity of PG~0101$+$039 assuming a canonical mass 
of 0.47\,M$_{\rm \odot}$ and tidally locked 
rotation. 
The light curve was then synthesised and was found to match the 
observed amplitude well.\\ 
{\it Results.} We verified that the light variations are due to ellipsoidal deformation and
that tidal synchronization is established for PG~0101$+$039. We conclude that 
this assumption should hold for all sdB binaries 
with orbital periods of less than half a day. Hence the masses can be derived from systems too faint to measure tiny light variations.

}

\maketitle

\section{Introduction \label{sec:intro}}

The masses of compact objects like white dwarfs, neutron stars and black 
holes are fundamental to astrophysics, but very difficult to measure. 
Close binary systems consisting of a visible primary and an invisible compact 
object are very useful to this end as the companion mass can be constrained 
from the radial velocity  and the light curve of the primary 
if the primary mass and 
the orbital inclination are
known. The latter 
can be measured in systems that are eclipsing or show tidally locked rotation.
If the mass of the primary can be estimated, the companion mass can then be 
derived.
The assumption of tidally locked rotation has often been used to determine the 
masses of neutron stars and black holes in 
known X-ray binaries (see Charles \& Coe \cite{charles} 
for a review). 
The same technique has recently been applied to KPD~1930$+$2752, a short period 
binary consisting of a subluminous B stars and a white dwarf
(Geier et al. \cite{geier1})\footnote{The companion is so massive that the system mass
may exceed the Chandrasekhar mass, making the system a viable Supernova Ia
progenitor candidate in the double degenerate scenario.}.  

sdB stars are core helium burning stars with
very thin hydrogen envelopes and masses around $0.5M_{\rm \odot}$ 
(Heber et al. \cite{heber1}). A large fraction of the sdB 
stars are members of short period binaries (Maxted et. al \cite{maxted2}; 
 Napiwotzki et al. \cite{napiwotzki6}). For these systems common envelope 
 ejection is the most probable formation channel (Han et al. \cite{han2}). 
 The companions of sdBs in these systems are predominantly white dwarfs 
 implying that the system has undergone two phases of common envelope ejection.




In order to establish bound rotation for an sdB star the synchronization time 
scale has to be smaller than its evolutionary life time 
($t_{\rm EHB}\approx10^{8}\,{\rm yrs}$).
Two theoretical concepts to compute synchronization times 
have been developed by Zahn (\cite{zahn}) and
Tassoul \& Tassoul (\cite{tassoul}), respectively.  
While the mechanism proposed by Zahn is not efficient enough to fully fit the observed levels of synchronization, the more efficient one of Tassoul \& Tassoul is matter of a controversy, given its free parameter dependence (see Claret et al. \cite{claret1}, \cite{claret2} and references therein).
Especially in the case of hot stars with radiative envelopes,
where tidal forces are less effective in synchronizing the stars, the
predictions of the two 
theoretical models at hand can differ by orders of magnitude. All prior studies were undertaken to match observations of hot main sequence stars. Hot subdwarf stars have similar temperatures as B-type main sequence stars, but are much smaller and the internal structure of these helium core burning objects is different. In addition, the fraction of sdBs residing in close binary systems is among the highest known of all types of stars. Observations of hot subdwarfs could provide a new benchmark to study the yet unresolved problem of tidal dissipation in radiative stellar envelopes.

Independent observational constraints are needed to prove or disprove synchronized rotation in hot subdwarf stars.
Ellipsoidal variations can be used to verify synchronization of the stellar surface because the 
light variations would then have to occur at exactly half the orbital period.
Two sdB + white dwarf binaries are known to  
show ellipsoidal variations at half of the 
orbital period (KPD\,0422+5421 Orosz \& Wade \cite{orosz}; KPD\,1930+2752, 
Geier et al. \cite{geier1}). 
However, both systems have short orbital periods of about $0.1\,{\rm d}$ and
theory predicts synchronization times much smaller the 
evolutionary time scale. 
As the synchronization time strongly increases with increasing period, 
we expect an upper limit to the period to exist at which the assumption of
tidally locked rotation breaks down.  
To this end it would be of utmost importance to find ellipsoidal variations in 
an sdB binary of longer period and to  
provide a stringent test for the theory of synchronization.
  
 Recently a suitable object has been found.
PG~0101$+$039, an sdB+WD binary (P=0.567 d, Maxted et al.
 \cite{maxted2}) was discovered to show very weak luminosity variation 
 at half the orbital period in a light curve in a 16.9 day long, almost
 uninterrupted light curve obtained with the MOST satellite (Randall et al. 
 \cite{randall}). 
 


In order to verify that we indeed see ellipsoidal
variations, we have to show that the observed light curve can be
consistently modelled. Beforehand, we have to derive 
the complete set of system parameters. As the spectrum is single 
lined, the
analysis of the radial velocity curve yields the mass function only.
Complementing it with an estimate of the sdB mass and with 
measurements of the sdB's
projected rotational velocity as well as its gravity allows to solve for all
binary parameters and compute the light curve. 

\section{Binary parameters \label{sec:par}}

\subsection{Radial velocity curve}
 
 Based on spectra obtained in 1998, Moran et al. (\cite{moran})
 determined 
the period  $P=0.569908\pm0.000007\,{\rm d}$. However, these ephemerides are 
not accurate enough to phase the MOST photometry because the time span of 
six years between spectroscopic and photometric observations is too long.
Therefore we combined the velocities of Moran (\cite{moran1})
with those from eight MMT-spectra taken in 1996, 1997 and 2002 
(Randall et al. \cite{randall}) and five spectra obtained with the Steward 2.3m Bok telescope from 2000.
The latter were determined using the double-precision version of the IRAF fxcor package,
against the combined template for the star. 
In addition we obtained three high resolution spectra ($R=30\,000$) in 2000 at the 2.2 m telescope at the German-Spanish Astronomical Center 
(DSAZ) equipped with the FOCES-spectrograph.  RVs were determined by cross 
correlation with a model spectrum at rest wavelength. 

This provided us with 57 velocities
distributed between 1996 and 2000. A sine curve was 
fitted to the observed velocities using an \(\chi^{2}\) minimising method and 
the power spectrum was generated (Napiwotzki et al. \cite{napiwotzki5}). 
The orbital parameters were measured: 
\(\gamma=7.3 \pm 0.2 {\rm \,km\,s^{-1}}\), 
\(K=104.5 \pm 0.3 \,{\rm km\,s^{-1}}\), 
\(P=0.569899 \pm 0.000001~ {\rm d}\) and \(HJD(0)=2452545.064275\). The period error
was derived with a 
bootstrapping algorithm. The period is sufficiently accurate to allow the MOST
photometry to be phased properly. 
However, it should be kept in mind that the error quoted is a purely statistical
one. To investigate possible systematic errors, which can occur when radial velocities 
obtained with different instruments are combined, we also used another method to verify our results.
In this case the fitting was performed with the system velocity $\gamma$ as additional free parameter
for every individual dataset. The results using both methods were perfectly consistent within 
the quoted errors. However, additional systematic effects may still be present. 

\subsection{Gravity and projected rotational velocity \label{sec:ana}}
 
Low resolution spectra obtained by Randall et al. (\cite{randall}) and 
Maxted et al. (\cite{maxted2}) were used to derived the atmospheric parameters.
Particular attention should be paid to the gravity determination as it provides 
a mass-radius-relation and its error propagates into the mass determination
(see. Sect. 4).    

Synthetic line profiles calculated
from metal line-blanketed LTE model atmospheres with solar metal content
(Heber et al. \cite{heber2}) were matched to the observed Balmer and helium 
line profiles
using a \(\chi^{2}\) fit procedure described by 
Napiwotzki et al. (\cite{napiwotzki1}). 
The resulting parameters are 
$T_{\rm eff}=27\,700\,{\rm K}$, $log\,g=5.55$, 
$log\,N(\rm He)/N(\rm H)=-2.62$ from the Randall et al. spectra and
 $T_{\rm eff}=27\,300\,{\rm K}$, $log\,g=5.50$, 
$log\,N(\rm He)/N(\rm H)=-2.71$ from the Maxted et al. spectra with formal  
statistical fitting errors of less than 100\,K, 0.02~dex and 0.02 dex,
respectively, which are unrealistically low. 
The true uncertainties are dominated by systematic inaccuracies in both the observations and 
model atmospheres and can be estimated 
from repeated observations and the use of different model grids. Taking into account the 
discussion of typical systematic errors applying this method in Geier et al.
(\cite{geier1}) we adopt $T_{\rm eff}=27\,500\pm 500\,{\rm K}$, $log\,g=5.53
\pm 0.07$, $log\,N(\rm He)/N(\rm H)=-2.66\pm 0.1$. 
  

In order to derive $v_{\rm rot}\,\sin{i}$ and the elemental abundances,
 we compared the observed high resolution spectra 
 with rotationally broadened, synthetic line profiles. 
 The projected rotational velocity was measured 
 simultaneously with the elemental abundances to 
 $v_{\rm rot}\sin{i}=10.9\pm1.1\,{\rm kms^{-1}}$ using 17 suitable metal 
 lines.

\begin{figure}[t!]
	\resizebox{8cm}{!}{\includegraphics{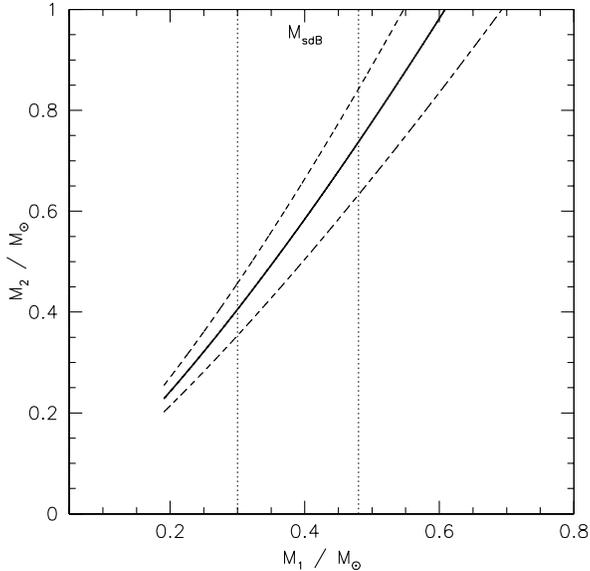}}
	\caption{Companion mass as a function of primary (sdB) mass. The dotted vertical lines mark the theoretical sdB mass range for the common envelope ejection channel (Han et al. \cite{han2}). The dashed lines mark the error limits of the companion mass.}
	\label{tlock}
\end{figure}

\subsection{Analysis \label{sec:analysis}}

The analysis strategy is the same as for KPD~1930$+$2752 and therefore is
described here only briefly. For details we refer
the reader to Geier et al. (\cite{geier1}).

Since the spectrum of PG~0101$+$039 is single-lined, it 
contains no information about the orbital motion of the companion, 
and thus only the mass function $f_{\rm m} = \frac{M_{\rm comp}^3 \sin^3i}{(M_{\rm comp} + M_{\rm sdB})^2} = \frac{P K^3}{2 \pi G}$ can be calculated. Although the RV semi-amplitude $K$ and the period $P$ are determined
 by the RV curve, $M_{\rm sdB}$, $M_{\rm comp}$ and $\sin^3i$ remain
 free parameters.


Nevertheless, the masses can be constrained by assuming tidal
synchronization. Combining the orbital parameters with an estimate of the sdB
mass and 
with the measurements of its $v_{\rm rot}\sin{i}$ and 
gravity allows the mass of the invisible companion to be constrained tightly. 
The mass of the sdB primary is constrained from the population synthesis models 
(Han et al. \cite{han2}) 
which predict a mass range of $M_{\rm sdB}$\,=\,0.30$-$0.48\,M$_{\rm \odot}$ 
for sdBs in binaries, that experienced a common envelope ejection. 
The mass distribution shows a sharp peak at a mass of 
about $0.47M_{\rm \odot}$. 

If the companion is synchronized the rotational velocity 
$v_{\rm rot}= \frac{2 \pi R_{\rm sdB}}{P}$ can be calculated. 
The radius of the primary is given by the mass radius relation 
$R = \sqrt{\frac{M_{\rm sdB}G}{g}}$. The measurement of the 
projected rotational velocity $v_{\rm rot}\,\sin\,i$ therefore allows us to 
constrain the inclination angle $i$.
With $M_{\rm sdB}$ as free parameter the mass function can be solved and the 
inclination angle as well as the companion mass can be derived. 
Because of $\sin{i} \leq 1$ a lower limit for the sdB mass is given 
by $M_{\rm sdB} \geq \frac{v_{\rm rotsini}^{2} P^{2}g}{4 \pi^{2}G}$.
There are no spectral signatures of the companion visible. 
A main sequence companion with a mass higher than 
$0.45M_{\rm \odot}$ can therefore be excluded because its 
luminosity would be sufficiently high to be detectable in the spectra.

The relation between the primary and the secondary mass is shown in 
Fig. \ref{tlock}. The allowed mass range for the companion is consistent 
with that of a white dwarf and therefore consistent with the common envelope 
ejection scenario. For the most likely sdB mass of $0.47M_{\rm \odot}$ the 
binary parameters are: $R_{\rm sdB}=0.19 \pm 0.02\,R_{\rm \odot}$, 
$M_{\rm WD}=0.72 \pm 0.10\,M_{\rm \odot}$, 
inclination $i=40 \pm 6\, {\rm ^{\circ}}$ and 
separation $a=3.1 \pm 0.4\,R_{\rm \odot}$.

\begin{figure}[t!]
	\resizebox{7.5cm}{!}{\includegraphics{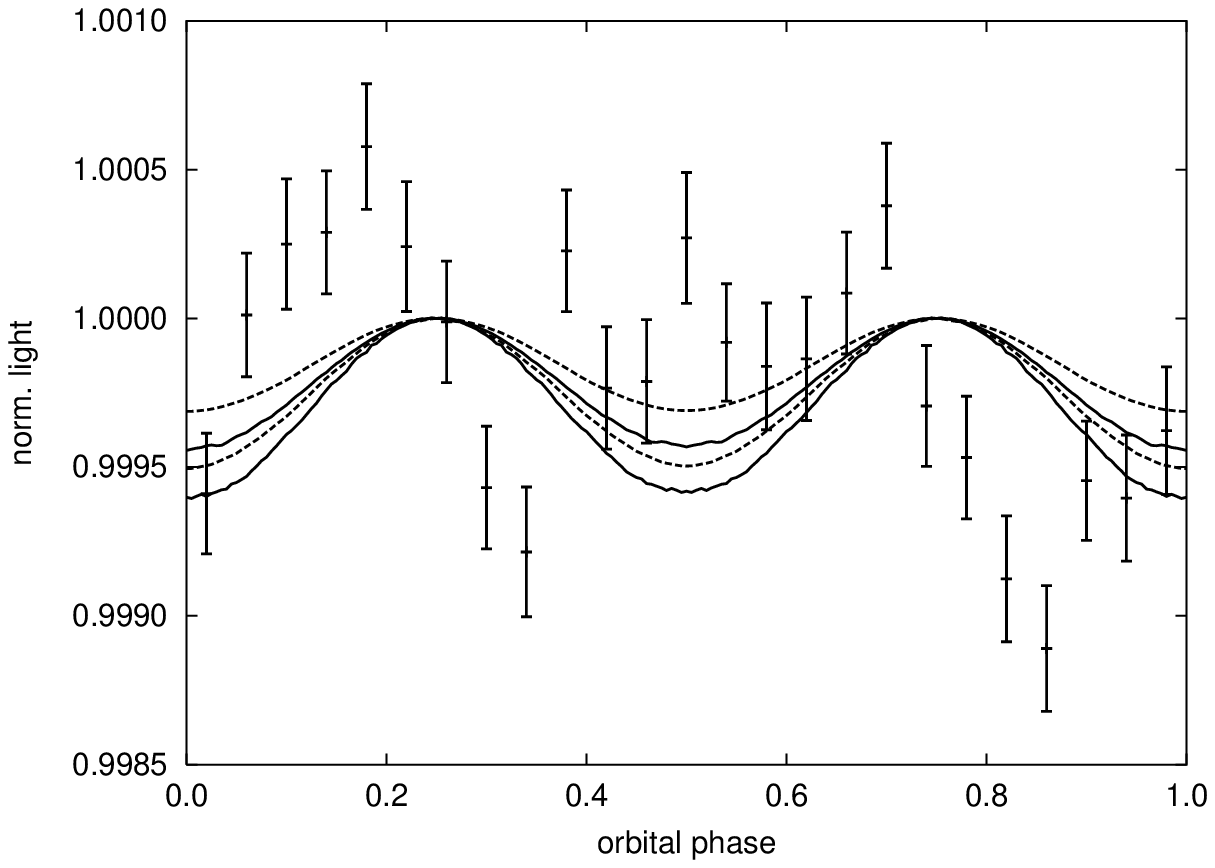}}
        \resizebox{7.5cm}{!}{\includegraphics{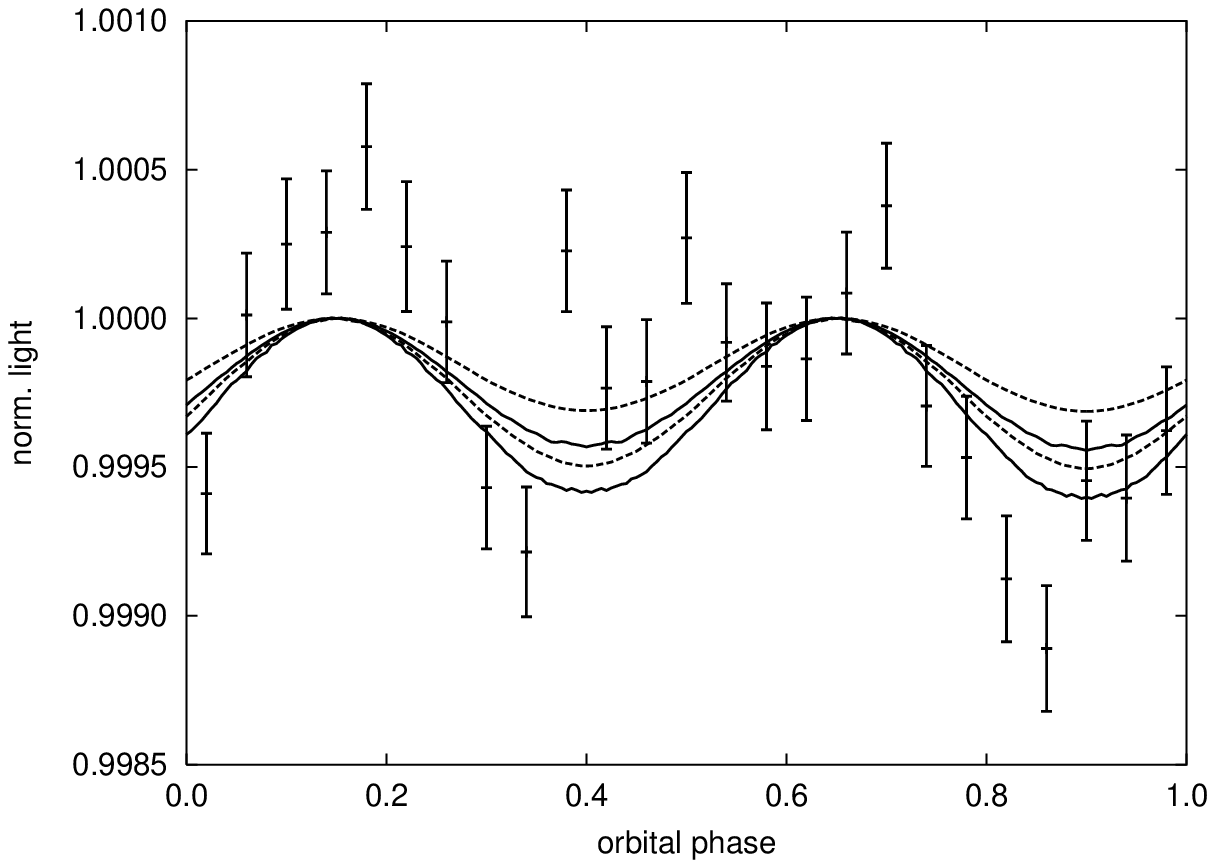}}
	\caption{Lightcurve data with superimposed models. 
The two solid curves confine the best fit models with parameters and associated uncertainties derived from 
spectroscopy under the assumption of orbital synchronization and $M_{\rm sdB}=0.3\,M_{\odot}$. The two dotted curves confine models for $M_{\rm sdB}=0.7\,M_{\odot}$. Phasing the data to the orbital solution results in a significant phase shift (upper panel). In the lower panel the model lightcurve is shifted by $-0.1$ in phase.}
	\label{ellips}
\end{figure}

\section{Light curve and elipsoidal variations}

The MOST photometric data were folded on the 
orbital period and re-binned to get a better signal to noise ratio (see Fig.
\ref{ellips}. 
As PG~0101$+$039 is a pulsating sdB star of V1093 Her type  
the light curve had to be pre-whitened for the 
pulsational frequencies beforehand. 


Each bin contains more than 400 original measurements. 
Light curve variations at half the orbital period with a semi-amplitude of 
$0.025\%$ were detected. Therefore  PG~0101$+$039 shows the smallest 
ellipsoidal variation ever measured. 
In order to compare with synthetic light curves we had to properly phase 
the photometry taking the zero point from the orbital solution. Because the amplitude of the variation is very low, Doppler boosting arising from the orbital motion affects the lightcurve significantly. A resulting factor of $(1-v(t)/c)$ was applied to the total flux to correct for this effect.

The light curve was modelled with the light curve synthesis and
solution code MORO based on the model by Wilson \& Devinney
(\cite{wilson}). The details of the Bamberg implementation are given by 
Drechsel et al. (\cite{drechsel}). The software uses a modified Roche model 
for light curve synthesis. It is capable of simulating the distortions of the stars induced
by a companion. Light curves for different component masses and orbital
inclinations were synthesised. For typical values of 
$M_{\rm sdB}$ appropriate ranges of $\sin{i}$ and
$R_{\rm sdB}$ values were computed as described in Sect. 
\ref{sec:par} covering the full parameter
space (including error limits) allowed by the spectroscopic analysis. 


We used sdB masses from $0.3-0.7M_{\rm \odot}$ for the model light 
curves and compared it to the observations in Fig. \ref{ellips}. The synthetic
light curve matches the semi-amplitude of the observed one quite well. 
Taking into account the extremely low amplitude of the 
variation, the consistency with the model is remarkable. 
However, there is a significant phase shift between the observed and the predicted 
light curve. The best fit to the data is phase shifted by $-0.1$ with respect 
to the model calculated with proper orbital phase (see Fig. \ref{ellips}). Despite our
efforts to derive a high precision orbital period, 
we can not rule out completely that this shift is caused by systematic effects 
when RV measurements from different instruments are combined.
To match the observed light curve would require the period to differ by
$0.00008\,{\rm d}$ from our results. Given the overall consistency of our orbital parameter determination, 
such a large deviation ($80\times$ period error) seems to be rather unlikely. 

%

\section{Discussion}

Tidally locked rotation in close binary systems has been assumed to measure 
masses of invisible compact companions, in particular in X-ray binaries. 
The synchronization time is very difficult to calculate for stars
with radiative envelopes and plagued with large uncertainties. 
Therefore, observational constraints are of utmost importance. 
Ellipsoidal variations can be used to verify the assumption at least for the surface layers. We applied this
technique to the sdB/WD binary PG~0101$+$039, 
for which a very weak luminosity 
variation at half the orbital period has been discovered in a light curve in a 
16.9 day long, 
almost uninterrupted light curve obtained with the MOST satellite.

From spectroscopy we measured the mass function, apparent rotation 
and surface gravity of PG~0101$+$039. Stellar evolution models suggest that
the sdB mass is close to 0.47\,M$_{\rm \odot}$. Assuming tidally locked 
rotation, this information is sufficient to solve for all parameters of the 
binary system. The companion mass is found to be  
$M_{\rm WD}=0.72 \pm 0.10\,M_{\rm \odot}$, typical for a white dwarf. 
The light curve was then synthesised and was found to match the 
observed amplitude well. However, a problem with the phasing of the light 
curve to the radial velocity curve became apparent. Due to a six year
difference between the MOST photometry and published radial velocities, 
the phase errors were far too large for any conclusion to be drawn. Therefore we
added 16 radial velocities from three observatories. The statistical error 
of the period decreased sufficiently to enable proper phasing of the 
photometry. The synthesised light curve was found to be offset by 0.1 cycles
from the observed one indicating that our systematic error estimate may be overly optimistic. 
Alternative explanations 
like supersynchronous rotation of the sdB that may cause the observed phase shift 
seem to be unlikely because a deviation of $10\%$ from equilibrium would require fast rotation 
of the sdB. In this case the inclination would be very low and the companion mass
would rise dramatically. 

A simultaneous measurement of the radial velocity curve and a high precision 
light curve would be necessary to solve this problem since the theoretical understanding of angular
momentum transfer in hot stars with radiative envelopes is still very limited.
In conclusion, we found strong indication that the surface rotation of the sdB star  
PG~0101$+$039 is tidally locked to its orbit.


The synchronization times for any given type of primary depend strongly on the 
orbital period (Zahn \cite{zahn}, Tassoul \& Tassoul \cite{tassoul}). Hence, other sdB stars in close binaries should also be synchronized if their orbital period is less than that of PG~0101$+$039
(P\,=\,$0.567\,{\rm d}$). 
Hence we conclude that tidally locked surface rotation is established in sdB binaries 
with orbital periods of less than half a day. Hence the assumption of tidally locked rotation can be safely applied to such systems, even if they are too faint to measure such extremely small light variations as observed here.

\begin{acknowledgements}

S.G. is supported by the Deutsche Forschungsgemeinschaft under grant He1354/40-3.
We would like to thank P. F. L. Maxted and T. Marsh for providing us with 
the radial velocity measurements of C. Moran. 
Furthermore we thank H. Drechsel for his kind advice.

\end{acknowledgements}

\end{document}